\documentclass[aps,prb,twocolumn,showpacs]{revtex4}

\usepackage{graphicx}

\begin{document}


\title{Doping effect of Cu and Ni impurities on the Fe-based superconductor Ba$_{0.6}$K$_{0.4}$Fe$_{2}$As$_{2}$}

\author{Peng Cheng$^1$, Bing Shen$^2$, Fei Han$^2$, and Hai-Hu
Wen$^3$$^{\star}$}

\affiliation{$^1$Department of Physics, Renmin University of
China, Beijing 100872, China}

\affiliation{$^2$National Laboratory for Superconductivity,
Institute of Physics and Beijing National Laboratory for Condensed
Matter Physics, Chinese Academy of Sciences, P.O. Box 603, Beijing
100190, China}

\affiliation{$^3$Center for Superconducting Physics and Materials,
National Laboratory of Solid State Microstructures and Department
of Physics, Nanjing University, Nanjing 210093, China}

\begin{abstract}
Copper and Nickel impurities have been doped into the iron
pnictide superconductor Ba$_{0.6}$K$_{0.4}$Fe$_{2}$As$_{2}$.
Resistivity measurements reveal that Cu and Ni impurities suppress
superconducting transition temperature T$_c$ with rates of $\Delta
T_c$/Cu-$1\%$ = -3.5 K and $\Delta T_c$/Ni-$1\%$ = -2.9 K
respectively. Temperature dependence of Hall coefficient R$_H$ of
these two series of samples show that both Cu-doping and Ni-doping
can introduce electrons into Ba$_{0.6}$K$_{0.4}$Fe$_{2}$As$_{2}$.
With more doping, the sign of R$_{H}$ gradually changes from
positive to negative, while the changing rate of Cu-doped samples
is much faster than that of Ni-doped ones. Combining with the
results of first-principles calculations published previously and
the non-monotonic evolution of the Hall coefficient in the low
temperature region, we argue that when more Cu impurities were
introduced into Ba$_{0.6}$K$_{0.4}$Fe$_{2}$As$_{2}$, the removal
of Fermi spectral weight in the hole-like Fermi surfaces is much
stronger than that in the electron-like Fermi surfaces, which is
equivalent to significant electron doping effect. DC magnetization
and the lattice constants analysis reveal that static magnetic
moments and notable lattice compression have been formed in
Cu-doped samples. It seems that the superconductivity can be
suppressed by the impurities disregard whether they are magnetic
or nonmagnetic in nature. This gives strong support to a pairing
gap with a sign reversal, like S$^\pm$. However, the relatively
slow suppression rates of T$_c$ show the robustness of
superconductivity of Ba$_{0.6}$K$_{0.4}$Fe$_{2}$As$_{2}$ against
impurities, implying that multi-pairing channels may exist in the
system.

\end{abstract}

 \maketitle

\section{Introduction}

The study of impurity effect is very important to the
understanding of superconductivity. It happens quite often that
the impurity induced suppression of superconductivity gives an
early hint of the unconventional pairing state.\cite{Review}
Conventional s-wave superconductors are sensitive to magnetic
impurities while robust to nonmagnetic impurities, which could be
explained by Anderson's theorem.\cite{AndersonTheorem} But for
cuprates with d-wave pairing symmetry, the superconductivity shows
little tolerance to both magnetic and nonmagnetic
impurities,\cite{xiaogang} nonmagnetic impurities could induce a
high density of states (DOS) due to the sign change of the gap on
a Fermi surface and cause rapid suppression of superconducting
transition temperature T$_c$. The study of the impurity effect on
superconductivity could give insights on the underlying paring
symmetry and superconducting mechanism. For example, rapid
suppression of the transition temperature T$_c$ in Al-doped
Su$_2$RuO$_{4}$ was the first indication that it is a novel
superconductor;\cite{Review} the observations of unusual charge
localization\cite{Ando} and enhanced antiferromagnetic (AFM)
correlations around Zn impurities\cite{NMRCuO} in
YBa$_{2}$Cu$_{3}$O$_{7-\delta}$ have spurred hot discussions of
the relationship between local AFM correlations and
superconductivity, which have important implications on the
superconducting mechanism in cuprate.

After the discovery of superconductivity in iron pnictides and chalcogenides\cite{Kamihara2008}, a lot of investigations show
unconventional superconducting mechanism and complicated gap
structures in these new
superconductors\cite{Mazin,Kuroki,neutron,Hanaguri,ZengBin,STMHHWen2012}.
Theoretically, an AFM fluctuation mediated fully-gapped
sign-reversal S$^\pm$ pairing state was
proposed\cite{Mazin,Kuroki,LeeDH,LiJX} and received supports from
the inelastic neutron scattering experiments\cite{neutron} and
scanning tunneling spectroscopy (STS)
measurements\cite{Hanaguri}. However there were also other
theoretical suggestions and experimental evidences for pairing
states ranging from S$^{++}$-wave to d-wave,or the existence of
gap
nodes.\cite{MuG,Hashimoto,LaFePO,Grafe,LuoXG,AokiPRB,Kontani,XueSTM,LiSY}.
Theoretical calculations show that the simple version of S$^\pm$ pairing state should be
fragile to impurities\cite{Parker,Onari,Bang,Kariyado}, only 1\%
impurity with moderate scattering potential could induce large in-gap state and
completely suppress superconductivity\cite{Onari}. Therefore plenty of
experiments are carried out on the impurity
induced suppression of superconductivity in Fe-based
superconductors\cite{Sato,LiYK,FeCuSeCava,
LiYK2,GuoYF,Chengpeng,Italy,GuoYF2,JLi2012}, in order to unravel the superconducting mechanism. Unfortunately the conclusions remain highly controversial. Further
experimental and theoretical works are clearly desired to clarify the
impurity effect in Fe-based superconductors, which could help us
better understand the pairing symmetry and superconducting
mechanism.

In this paper, we report the doping effect of Cu and Ni impurities
on the superconductor Ba$_{0.6}$K$_{0.4}$Fe$_{2}$As$_{2}$. Both of
these impurities could suppress superconductivity in certain rates
and cause electron-doping effects.  According to many
other first-principles-calculations, although the dopant Cu in Fe-based
superconductors is in the valence state of +1 and with a fully
occupied $d$ orbit, the doping of Cu seems to introduce more
electrons than doping Ni as revealed from the Hall effect
measurements. DC magnetization measurements clearly show that the Cu-doping in Ba$_{0.6}$K$_{0.4}$Fe$_{2}$As$_{2}$ can induce the magnetic impurities, while the suppression to superconductivity is similar to the Ni-doping, which yields non- or weak magnetic impurity centers. Our results clearly indicate that the superconductivity can be suppressed by impurities at the Fe sites, disregard whether they are magnetic or nonmagnetic in nature.

\section{Experimental results}

The Cu-doped and Ni-doped polycrystalline samples
Ba$_{0.6}$K$_{0.4}$(Fe$_{1-x}$TM$_{x}$)$_{2}$As$_{2}$ (TM = Cu and
Ni) were fabricated by solid state reaction method, the specific
fabrication process was described in our previous
paper\cite{Chengpeng}. The x-ray diffraction (XRD) measurement was
performed using an MXP18A-HF-type diffractometer with
Cu-K$_{\alpha}$ radiation. The analysis of x-ray diffraction data
was done by using the softwares POWDER-X and Fullprof, the obtained
results are consistent with each other. The AC susceptibility
measurements were carried out through an Oxford cryogenic system
Maglab-EXA-12. The resistivity, magnetoresistance and Hall effect
were measured with a Quantum Design instrument physical property
measurement system (PPMS), and the DC magnetization by a Quantum
Design instrument SQUID (MPMS-7).

The temperature dependence of resistivity of samples
Ba$_{0.6}$K$_{0.4}$(Fe$_{1-x}$TM$_{x}$)$_{2}$As$_{2}$ (TM=Cu,Ni)
were shown in Fig.1(a) and Fig.1(b) respectively. The T$_c$ in
impurity-free sample Ba$_{0.6}$K$_{0.4}$Fe$_{2}$As$_{2}$ is about
38 K, which is obviously an optimally hole-doped sample. As we can
see the superconducting transition temperature was gradually
suppressed and the residual resistivity (RR) rose upon the doping
of either Cu or Ni impurities. The specific values of changing
rates for T$_c$ and RR were calculated and presented later. We
also notice that the values of normal state resistivity at above
150 K tend to increase with Cu doping while decrease with more Ni
doping. This difference could be attributed to the different
change of electronic band structures brought by these two
different impurities, which would be also discussed in next
section.

\begin{figure}
\includegraphics[width=8cm]{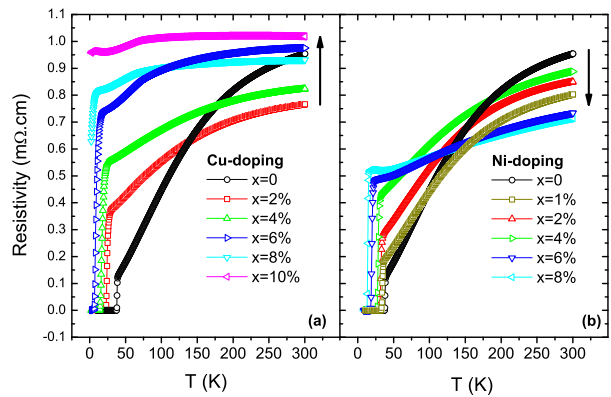}
\caption {(color online) (a) Temperature dependence of resistivity
of the Ba$_{0.6}$K$_{0.4}$(Fe$_{1-x}$Cu$_{x}$)$_{2}$As$_{2}$
samples under zero field.  (b) Temperature dependence of
resistivity of the
Ba$_{0.6}$K$_{0.4}$(Fe$_{1-x}$Ni$_{x}$)$_{2}$As$_{2}$ samples
under zero field. With the doping of more Ni impurities, both the
suppression rate of T$_c$ and the increasing rate of residual
resistivity are relatively slow compared with that of Cu-doped
samples. } \label{Fig1}
\end{figure}

\begin{figure}
\includegraphics[width=8cm]{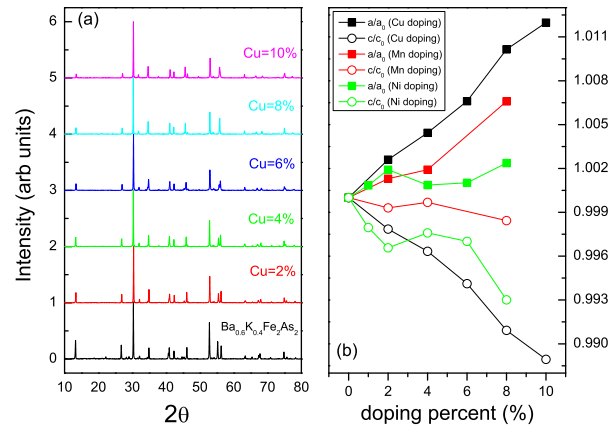}
\caption {(color online) (a)The x-ray diffraction patterns for
Ba$_{0.6}$K$_{0.4}$(Fe$_{1-x}$Cu$_{x}$)$_{2}$As$_{2}$, the samples
are quite clean and no obvious phase segregation can be detected.
(b) Evolution of a-axis and c-axis lattice constants with the
doping of Cu, Ni and Mn impurities respectively. All the data are
normalized by the values of undoped sample. The data points of
Mn-doped samples are derived from our previous
report\cite{Chengpeng}. } \label{Fig2}
\end{figure}

The x-ray diffraction patterns of Cu-doped samples were shown in
Fig.2(a), the rather pure single phase indicate good quality of
our samples. Fig.2(b) displays the evolution of lattice parameters
for different doping samples. The values of Mn-doped ones were
taken from our previous paper\cite{Chengpeng}. Upon doping
transition metal impurities onto the Fe sites, the lattice
parameters for a-axis increase while that for c-axis decrease,
which indicates that the crystal lattice is gradually compressed
along the c-axis direction with impurity-doping. One can see that
although the overall changing trend for the three different series
of samples are similar, the Cu-doped samples show much stronger
lattice compression compared to Ni and Mn doped ones.

\begin{figure}
\includegraphics[width=8cm]{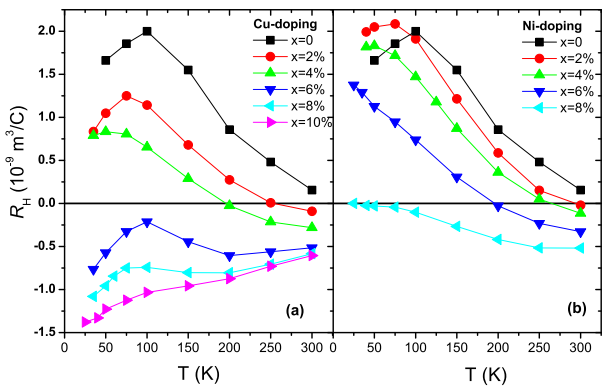}
\caption {(color online) Temperature dependence of Hall
coefficient R$_H$ = $\rho_{xy}$/H measured at 9 T for (a)
Ba$_{0.6}$K$_{0.4}$(Fe$_{1-x}$Cu$_{x}$)$_{2}$As$_{2}$ and (b)
Ba$_{0.6}$K$_{0.4}$(Fe$_{1-x}$Ni$_{x}$)$_{2}$As$_{2}$. The
sign-change of R$_H$ with doping is obvious in the two sets of
samples.} \label{Fig3}
\end{figure}

In Fig.3(a) and Fig.3(b), we show the temperature dependence of
Hall coefficients R$_H$ for
Ba$_{0.6}$K$_{0.4}$(Fe$_{1-x}$TM$_{x}$)$_{2}$As$_{2}$ (TM=Cu,Ni)
respectively. Hall coefficients gradually change from positive
values to negative values as either Cu or Ni doping. This
indicates that both Cu and Ni introduced electrons into the
hole-doped superconductor Ba$_{0.6}$K$_{0.4}$Fe$_{2}$As$_{2}$. We
also noticed the changing rate of $R_H$ is higher for Cu-doped
ones. For Cu-doped samples, the values of R$_H$ become completely
negative for x=6\% in the whole measured temperature region, while
for Ni-doping samples, R$_H$ become completely negative at the
doping level of x=8\%. From the temperature dependence of the Hall
coefficient R$_H$ for the slightly doped samples, one can see a
non-monotonic temperature dependence in the low temperature
region. While it becomes more monotonic like when the
electron-like charge carriers dominates. The doping effect of
R$_H$ cannot be easily understood within the rigid band model. In
Fig.4(a) and 4(b) we present the doping dependence of the Hall
coefficient R$_H$ measured at 200 K. It is clear that the
effective doping of electrons by adding Cu is stronger than that
by adding Ni. A simple normalization of the Hall efficient suggest
that the rigid model seems working at high temperatures, that is,
doped electrons ratio Cu(electrons)/Ni (electrons)=3/2, this is
very much like that of the Co, and Ni doping\cite{Canfield1}. This
interesting observation should be reconciled with the calculations
of electronic structures.

\section{Analysis and discussions}
\subsection{Magnetic moments introduced by doping Cu ions in Ba$_{0.6}$K$_{0.4}$Fe$_{2}$As$_{2}$}
In order to investigate the impurity effect in
Ba$_{0.6}$K$_{0.4}$Fe$_{2}$As$_{2}$, we measured the magnetization
of the doped samples in a magnetic field of 1 T. The data of
Cu-doped samples were shown in Fig.5(a). The magnetization curve
exhibits a typical T-linear behavior for undoped
Ba$_{0.6}$K$_{0.4}$Fe$_{2}$As$_{2}$ sample in the high temperature
region. This linear behavior was interpreted as the origin of the
short range AF correlation\cite{ZhangGM}. However with more
Cu-doping, a Curie-Weiss like behavior emerges and gets more and
more strong. For all Cu-doped samples, the data of magnetic
susceptibility at the temperature below 150 K could be well fitted
by the Curie-Weiss law. Based on the fitting results, we
calculated the average magnetic moments of one
Ba$_{0.6}$K$_{0.4}$(Fe$_{1-x}$Cu$_{x}$)$_{2}$As$_{2}$ molecule,
which is presented in Fig.5(b). It is clear that the doping of Cu
gradually introduces magnetic moments into this system and the
magnetic moment is getting stronger with more Cu doping. In the
inset of Fig.5(b), the fitted curve of Cu-doped 8\% sample was
shown as an example of the fitting. Fe-based superconductors are
in proximity to magnetism, so the disruption of the electronic
structure by scattering would be expected to lead the formation of
local moments around Cu sites\cite{FeCuSeCalc}. Through the
fitting results of the M-T curve with Curie-Weiss law:
\begin{equation}
\chi=\chi_0+\frac{C}{T+T_N}\\
\end{equation}
where $C=\mu_0\mu_J^2/3k_B$, we can get the magnetic moment
$\mu_J$ for each
Ba$_{0.6}$K$_{0.4}$(Fe$_{1-x}$Cu$_{x}$)$_{2}$As$_{2}$ molecule,
which is shown in Fig.5 (b). The values of magnetic moments
gradually increase with more Cu-doping and seem to get saturated
at the doping level of 6\%. Because of the $d^{10}$ configuration
of Cu, as claimed by the first-principles calculations, it is
quite difficult to understand why a magnetic moment is induced at
the Cu-sites. One possible picture would be that these magnetic
moments may not exist at the Cu site, but could be distributed
over the Fe ions around the Cu site, like the case in the cuprate
superconductors near a Zn impurity.

\begin{figure}
\includegraphics[width=8cm]{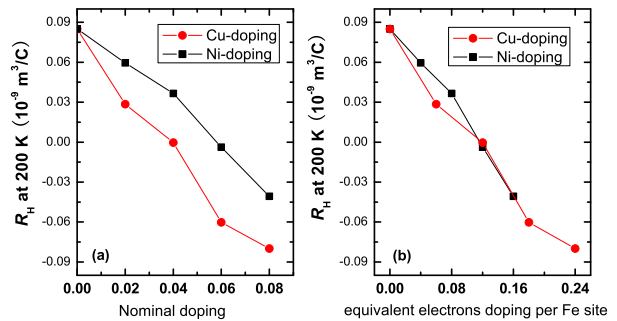}
\caption {(color online) (a) Doping dependence of $R_H$ at the
temperature of 200 K for
Ba$_{0.6}$K$_{0.4}$(Fe$_{1-x}$Cu$_{x}$)$_{2}$As$_{2}$ and
Ba$_{0.6}$K$_{0.4}$(Fe$_{1-x}$Ni$_{x}$)$_{2}$As$_{2}$ samples.
(b)$R_H$ at 200 K as a function of equivalent electrons doping per
Fe site. For Cu-doped samples we multiply three and for Ni-doped
ones we multiply two.} \label{Fig4}
\end{figure}

\subsection{Suppression to T$_c$ by the impurities}

From the x-ray data, we can see that transition metals such as Cu,
Ni and Mn could be easily doped on the Fe sites in Fe-based
superconductor Ba$_{0.6}$K$_{0.4}$Fe$_{2}$As$_{2}$. The
substitution of these impurities could all suppress
superconductivity and raise the values of residual resistivity
(RR), which means that these transition metals act as scattering
centers. We should mention that, although the doping can induce
the partial charge doping to the system, while the suppression to
superconductivity here is mainly induced by the impurity
scattering. This can be corroborated by the simple linear relation
between the residual resistivity and the normalized T$_c$
suppression as shown in Fig.1, Fig.6(b) and Fig.7. According to
the Abrikosov-Gorkov formula,\cite{AG} if the impurities act as
strong pair breakers, the T$_c$ suppression due to pair breaking
is essentially related to the impurity scattering rate $k_B\Delta
T_c\approx \pi\hbar/8\tau_{imp}\propto\rho_0$, where $\rho_0$ is
the residual resistivity. In Fig.6(a), the doping dependence of
T$_c$-suppression rates for Cu, Ni and Mn were presented, T$_c$
decreases almost linearly with increasing the nominal impurity
doping level. Through the calculation of the slopes, we can get
the average change of T$_c$ values per doping percent: $\Delta
T_c$/Mn-$1\%$ = -4.2 K, $\Delta T_c$/Cu-$1\%$ = -3.5 K, $\Delta
T_c$/Ni-$1\%$ = -2.9 K. While in Fig.6(b), we show the
corresponding doping dependence of residual resistivity. Similarly
we got $\Delta \rho_0$/Mn-$1\%$ = 0.107 m$\Omega$ cm, $\Delta
\rho_0$/Cu-$1\%$ = 0.093 m$\Omega$ cm, $\Delta \rho_0$/Ni-$1\%$ =
0.071 m$\Omega$ cm. From these data, it is clear that the
impurity-doped samples which have higher rising rate of RR could
cause a more rapid T$_c$ suppression. Thus Mn and Cu are
relatively strong scattering centers, while Ni causes relatively
weak impurity scattering effect.

From the lattice parameters point of view, as displayed in
Fig.2(b), one can see that the changing tendency of lattice
constants upon impurities is almost the same for all three series
of samples, we also noticed the values of a-axis and c-axis in
Cu-doped samples change faster than all the other impurity-doped
samples. This means that the crystal lattices of Cu-doped samples
are most strongly compressed and suffer stronger lattice change
than the others.

\begin{figure}
\includegraphics[width=8cm]{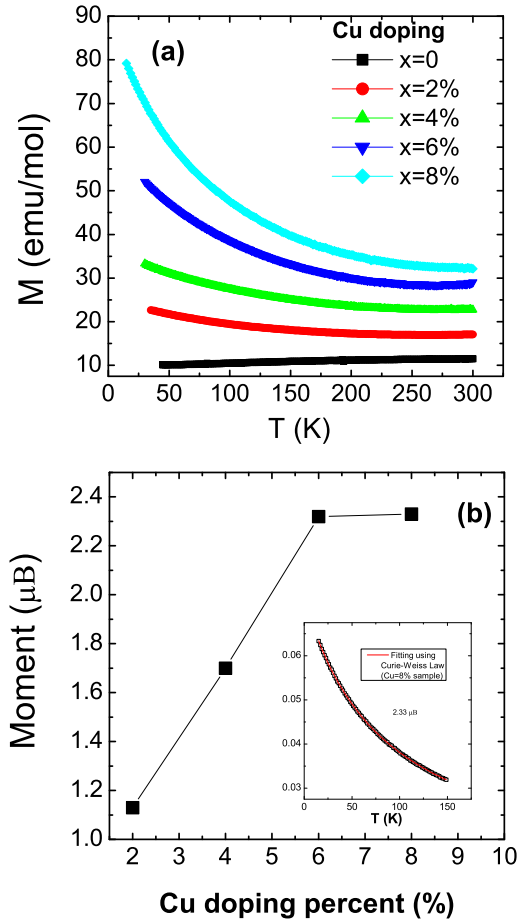}
\caption {(color online)(a) DC magnetic susceptibility as a
function of temperature of
Ba$_{0.6}$K$_{0.4}$(Fe$_{1-x}$Cu$_{x}$)$_{2}$As$_{2}$ samples. A
DC magnetic field of 1 T was applied in the measurement. (b) The
average magnetic moments of one
Ba$_{0.6}$K$_{0.4}$(Fe$_{1-x}$Cu$_{x}$)$_{2}$As$_{2}$ molecule
were calculated according to the results of Curie-Weiss fitting.
The inset shows the fitting result of M-T curve in low temperature
region of sample Cu=8\% using the Curie-Weiss law.} \label{Fig5}
\end{figure}

It has been realized earlier that the main features of band
structure and electronic density of states (DOS) remain the same
after either cobalt or nickel was doped in the iron-site in
iron-pnictides\cite{XuG}. A lot of researches also point out that,
in the parent compound of Fe-based superconductors, the
substitution of one Fe ion with one Co ion means the doping of one
more electron, while doping one Ni ion can provide two
electrons\cite{Hosono,Canfield1,3dcalcKuwei}. A simple rigid band
model could explain the experimental results. Since Cu is the
element behind Ni in the periodical table and has one more $d$
electron than Ni, a straightforward thinking is that the doping of
Cu means adding three more electrons. From the Hall data in Fig.3,
people could easily have the first feeling that the values of
$R_H$ in Cu-doped samples change to be negative more quickly than
that in Ni-doped samples. This trend is further illustrated in
Fig.4(b), where we plot the $R_H$ values at the temperature of 200
K for the two series of samples. One can also see that the
evolution of $R_H$ with doping could be scaled together for Cu and
Ni doped samples if we assume the equivalent doping of one Cu is
adding three electrons while two electrons for Ni. The similar
data and interpretation have also been reported by Canfield et
al\cite{Canfield1}. However according to the results of many first
principles calculations of the electronic density of states in
Cu-doped iron pnictides and chalcogenides
materials\cite{FeCuSeCalc,3dcalc,canfieldcalc}, unlike the case of
Co-doping or Ni-doping, Cu exhibits split-band behavior with the
$d$-electrons situated well below (3 to 4 eV) $E_F$, the $d$ bands
of Cu are fully occupied for a nominal $d^{10}$ configuration.
Therefore a state of Cu$^{+}$ would be expected, which indicates
effective hole doping through the substitution of Fe$^{2+}$ by
Cu$^+$. However from the shifts of the Fe density of states after
Cu doping and the Hall data in this paper, an electron doping
picture was suggested. This controversial situation could be
reconciled by the analysis of the Fermi surfaces evolution. Chadov
et al\cite{FeCuSeCalc} calculated the Fermi surfaces of
Fe$_{1-x}$Cu$_x$Se, where they show that Cu-doping is highly
disruptive to the electronic structure of FeSe superconductor and
caused strong loss of Fermi spectral weight for the hole-like
Fermi surfaces. Based on this picture, we could provide a possible
explanation for the effective strong electron doping effect caused
by Cu: As more Cu was doped into the lattice, both the hole
pockets at around $\Gamma$ point and the electron pockets at
around M point are gradually destroyed. This loss of Fermi
spectral weight is especially rapid and stronger for the hole
pockets, which could result in an equivalent electron doping
effect for Cu-doped Ba$_{0.6}$K$_{0.4}$Fe$_{2}$As$_{2}$. However,
the remaining controversy is that why in the high temperature
region, the simple rigid model seems working very well, as
displayed in Fig.4(b).

Considering the strong impurity scattering effect brought by Cu,
the disorder induced localization is usually expected. Through the
comparison of the normal state resistivity between Cu-doped and
Ni-doped samples, we noticed that with more impurities, the normal
state resistivity at high temperature region (150 K to 300 K,
approximately) go higher with Cu-doping and have a trend of going
down with Ni-doping. This phenomenon is consistent with the above
discussions, namely the doping of Ni leads to the formation of a
coherent electronic structure with the original
Ba$_{0.6}$K$_{0.4}$Fe$_{2}$As$_{2}$, while the doping the Cu is
destructive to the original electronic structure and causes
localization effect. However, we should emphasize that, the
suppression to the superconductivity is not induced by the
localization of the electrons, but rather by the impurity
scattering. This statement is specially valid below a moderate
doping level. Actually in our data shown in Fig.1(a), we cannot
see the up-turn of resistivity in the low temperature region,
which should be the case if the localization would behave an
important role here.

\begin{figure}
\includegraphics[width=8cm]{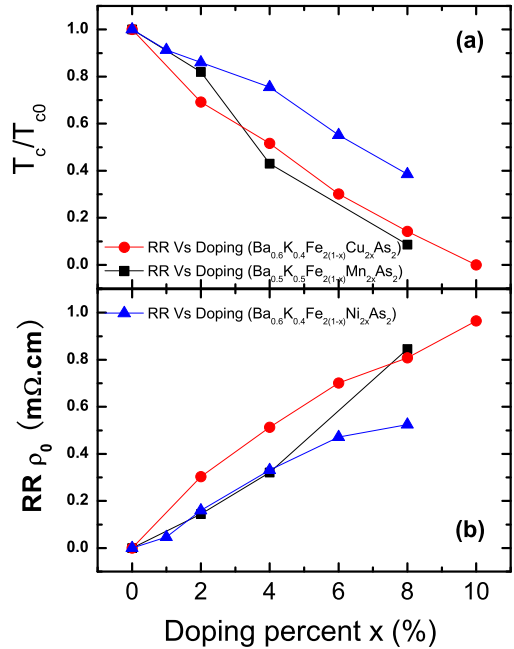}
\caption {(color online) (a) Doping dependence of T$_c$/T$_c$$_0$
in Cu-, Ni- and Mn-doped samples. The values of T$_c$ are
determined from the resistivity data and taking the 90\%$\rho_n$
criteria. (b) Doping dependence of residual resistivity $\rho_0$
in Cu-, Ni- and Mn-doped samples. $\rho_0$ is determined through
the extrapolation of the normal state data to T = 0 K. The data of
Mn-doped samples was taken from our previous
paper\cite{Chengpeng}. } \label{Fig6}
\end{figure}

It has been widely discussed, in the only two high-$T_c$
superconductor families discovered by now, the cuprates and
iron-pnictides have many similarities, for example, their parent
compounds are both AFM ordered. However, they also have a lot of
differences. For the impurity doping effect, we would say, these
two families act quite differently. For
Ba$_{0.6}$K$_{0.4}$Fe$_{2}$As$_{2}$, we choose a cuprate analog
for comparison, namely La$_{1.85}$Sr$_{0.15}$CuO$_{4}$. These two
superconductors are both hole doped and with almost the same $T_c$
of about 38 K\cite{xiaogang}. But for the latter, a d-wave cuprate
superconductor, the superconductivity can be killed completely
with a slight doping of impurities, for example, only about 3\% of
Zn doping can diminish the superconductivity \cite{xiaogang},
which yields a suppression rate of $\Delta T_c$/$1\%impurity$ =
-11 K. This is of course in sharp contrast with the
impurity-doping in Ba$_{0.6}$K$_{0.4}$Fe$_{2}$As$_{2}$, which is
usually below $\Delta T_c$/$1\%impurity$ = -4.2 K.
\cite{Chengpeng} Especially for the case of Cu-doping in this
paper, the electron-doping effect itself may result in a partial
T$_c$-suppression, therefore the T$_c$-suppression caused by
impurity scattering of Cu should be considered as moderate. The
different responses to impurities for these two kinds of
superconductors indicate that they could have different pairing
and superconducting mechanism. One explanation would be that the
pairing in the iron pnictide is not induced solely by the simple
inter-pocket pair-scattering. The intra-pocket scattering may also
contribute significantly to the pairing
strength.\cite{HirschfeldReview} We also noticed there were
reports about more rapid suppression of superconductivity in
Zn-doped LaFeAsO$_{0.85}$\cite{GuoYF}. This means that the
impurity effect in iron-pnictides is complicated, perhaps it is
different for different compounds and doping regions\cite{LiYK2}.
Theoretically both the S$^\pm$ and d-wave pairing are fragile to
impurities, therefore Kontani et al. proposed an s-wave
superconducting state without sign reversal (namely the
$S_{++}$-wave state) for Fe-based superconductors based on
d-orbital fluctuations considering the robustness
superconductivity against impurities\cite{Kontani}. In Fig.7, we
present the correlation between T$_c$ and residual resistivity
(the scattering rate). One can see that the suppression to T$_c$
is more or less the same. Concerning the different magnetic
moments induced by these dopants, our experiment indicates that
the superconductivity can be suppressed by impurities disregard
whether they are magnetic or nonmagnetic in nature. This
phenomenon can certainly place more weight on the side of S$^\pm$
pairing. Further research is highly desired to address why the
suppression rate to superconductivity by the impurities in the
iron pnictide superconductors is moderate, not as strong as
expected by a simplified model of the S$^\pm$ pairing.

\begin{figure}
\centering
\includegraphics[bb=0 0 380 300,width=8cm]{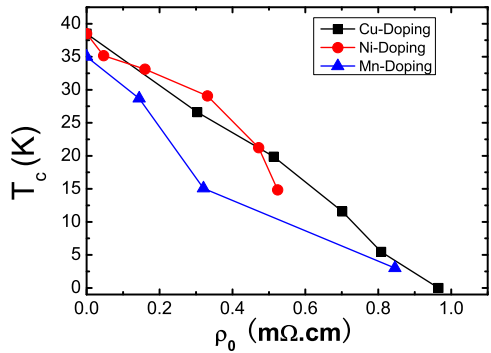}
\caption {(color online) Correlations between the superconducting
transition temperature and the residual resistivity in the Mn, Ni
and Cu doped samples. The experiment for Mn doping was done in the
system of Ba$_{0.5}$K$_{0.5}$Fe$_2$As$_2$. That for Ni and Cu
doping was done in the system of Ba$_{0.6}$K$_{0.4}$Fe$_2$As$_2$.}
\label{Fig7}
\end{figure}

\section{Concluding remarks}
In summary, the impurity doping effects of Cu and Ni on the
structure, transport properties, magnetism and superconductivity
of Ba$_{0.6}$K$_{0.4}$Fe$_{2}$As$_{2}$ have been studied. The
substitution of these impurities on the Fe sites could all
suppress T$_c$ and raise the values of residual resistivity in
certain rates. The measurements of Hall coefficients show more
electron-like charge carriers have been introduced into the system
for Cu-doped samples than Ni-doped ones. This significant electron
doping effect in Cu-doped samples could  be explained possibly by
the relatively stronger loss of Fermi spectral weight in the
hole-like Fermi surfaces than the electron-like ones in
Ba$_{0.6}$K$_{0.4}$(Fe$_{1-x}$Cu$_{x}$)$_{2}$As$_{2}$. Although
the theoretical calculations suggest that the $d$ bands of Cu are
fully occupied for a nominal $d^{10}$ configuration, magnetic
moments with moderate strength have been found the in Cu-doped
samples. The common behavior among the samples by doping Mn, Ni
and Cu strong suggest that the superconductivity can be suppressed
by impurities disregard whether they are magnetic or nonmagnetic
in nature. This gives of course more weight to the S$^\pm$
pairing. However, the suppression rate to T$_c$ is moderate, not
as strong as expected by the simple model of S$^\pm$ pairing. We
attribute this to the multiple pairing channels, such as the
inter-pocket or intra-pocket pairing in the samples.

\begin{acknowledgments}
We appreciate the useful discussions with D. J. Singh, I. I.
Mazin, P. Hirschfeld, D.-H. Lee, Q. H. Wang and G.-M. Zhang. This
work is supported by the NSFC of China, the Ministry of Science
and Technology of China (973 Projects: 2011CBA001002,
2012CB821403, 2010CB923002). The work at RUC is supported by NSFC
(No.11204373) and the Fundamental Research Funds for the central
Universities, and the Research Funds of Renmin University of
China.
\end{acknowledgments}

$^{\star}$ hhwen@nju.edu.cn


\begin{thebibliography}{99}
\bibitem{Review} A. V. Balatsky, I. Vekhter, and J. X. Zhu, Rev. Mod. Phys. {\bf78}, 000373 (2006).
\bibitem{AndersonTheorem} P. W. Anderson, J. Phys. Chem. Solids {\bf11}, 26 (1959).
\bibitem{xiaogang}G. Xiao, M. Z. Cieplak, J. Q. Xiao, and C. L. Chien, Phys. Rev. B{\bf42}, 8752 (1990).
\bibitem{Ando} K. Segawa, and Y. Ando, Phys. Rev. B {\bf59}, R3948 (1999).
\bibitem{NMRCuO} M. H. Julien, T. Feher, M. Horvatic, C. Berthier, O. N. Bakharev, P. Segransan, G. Collin, and J. F. Marucco, Phys. Rev. Lett. {\bf84}, 3422 (2000).
\bibitem{Kamihara2008} Y. Kamihara, T. Watanabe, M. Hirano, and H. Hosono, J. Am. Chem. Soc. {\bf130}, 3296 (2008).
\bibitem{Mazin} I. I. Mazin, D. J. Singh, M. D. Johannes, and M. H. Du, Phys. Rev. Lett. {\bf101}, 057003 (2008).
\bibitem{Kuroki} K. Kuroki, S. Onari, R. Arita, H. Usui, Y. Tanaka, H. Kontani, and H. Aoki, Phys. Rev. Lett. {\bf101}, 087004 (2008).
\bibitem{neutron} A. D. Christianson, E. A. Goremychkin, R.
Osborn, S. Rosenkranz, M. D. Lumsden, C. D. Malliakas, I. S.
Todorov, H. Claus, D. Y. Chung, M. G. Kanatzidis, R. I. Bewley,
and T. Guidi, Nature {\bf456}, 930 (2008).
\bibitem{Hanaguri} T. Hanaguri, S. Niitaka, K. Kuroki, and H. Takagi, Science {\bf328}, 474 (2010).
\bibitem{ZengBin} B. Zeng, G. Mu, H. Q. Luo, T. Xiang, I. I.
Mazin, H. Yang, L. Shan, C. Ren, P. C. Dai, and H. H. Wen, Nat.
Commun. {\bf1}, 112 (2010).
\bibitem{STMHHWen2012}Z. Y. Wang, H. Yang, D. L. Fang, B. Shen, Q.
H. Wang, L. Shan, C. L. Zhang, P. C. Dai, and H. H. Wen, Nature
Physics {\bf9}, 42 (2013).
\bibitem{LeeDH} F. Wang, H. Zhai, Y. Ran, A. Vishwanath, and D. H. Lee, Phys. Rev. Lett. {\bf102}, 047005 (2009).
\bibitem{LiJX}Z. J. Yao, J. X. Li, and Z. D. Wang, New J. Phys. {\bf11}, 025009 (2009).



\bibitem{MuG} G. Mu, X. Y. Zhu, L. Fang, L. Shan, C. Ren, and H. H. Wen, Chin. Phys. Lett. {\bf25}, 2221 (2008). G. Mu, H. Q. Luo, Z. S. Wang, L. Shan, C. Ren, and H. H. Wen, Phys. Rev. B {\bf79}, 174501 (2009).
\bibitem{Hashimoto}K. Hashimoto, T. Shibauchi, T. Kato, K. Ikada, R. Okazaki, H. Shishido, M. Ishikado, H. Kito, A. Lyo, H. Eisaki, S. Shamoto, and Y. Matsuda, Phys. Rev. Lett. {\bf102}, 017002 (2009).
\bibitem{LaFePO} J. D. Fletcher, A. Serafin, L. Malone, J. Analytis, J-H Chu, A. S. Erickson, I. R. Fisher, A. Carrington. Phys. Rev. Lett. {\bf102}, 147001 (2009).
\bibitem{Grafe}H.-J. Grafe, D. Paar, G. Lang, N. J. Curro, G. Behr, J. Werner, J. H. Borrero, C. Hess, N. Leps, R. Klingeler, and B. Buchner, Phys. Rev. Lett. {\bf101}, 047003(2009).
\bibitem{LuoXG}X. G. Luo, M. A. Tanatar, J. P. Reid, H. Shakeripour, N. Doiron-Leyraud, N. Ni, S. L. Budko, P. C. Canfield, H. Q. Luo, Z. S. Wang, H. H. Wen, R. Prozorov, and L. Taillefer, Phys. Rev. B {\bf 80}, 140503(R) (2009).
\bibitem{AokiPRB}K. Kuroki, H. Usui, S. Onari, R. Arita, and H. Aoki, Phys. Rev. B {\bf79}, 224511 (2009).
\bibitem{Kontani}H. Kontani, and S. Onari, Phys. Rev. Lett. {\bf104}, 157001 (2010).
\bibitem{XueSTM}C. L. Song, Y. L. Wang, P. Cheng, Y. P. Jiang, W.
Li, T. Zhang, Z. Li, K. He, L. L. Wang, J. F. Jia, H. H. Hung, C.
W, X. C. Ma, X. Chen, and Q. K. Xue, Science {\bf332}, 1410
(2011).
\bibitem{LiSY}J. K. Dong, S. Y. Zhou, T. Y. Guan, H. Zhang, Y. F.
Dai, X. Qiu, X. F. Wang, Y. He, X. H. Chen, and S. Y. Li, Phys.
Rev. Lett. {\bf104}, 087005 (2010).

\bibitem{Parker} D. Parker, O. V. Dolgov, M. M. Korshunov, A. A. Golubov, and I. I. Mazin, Phys. Rev. B {\bf78}, 134524 (2008).
\bibitem{Onari} S. Onari, and H. Kontani, Phys. Rev. Lett. {\bf103}, 177001 (2009).
\bibitem{Bang} Y. Bang, H. Choi, and H. Won, Phys. Rev. B {\bf79}, 054529 (2009).
\bibitem{Kariyado} T. Kariyado, and M. Ogata, J. Phys. Soc. Jpn. {\bf79}, 083704 (2010).




\bibitem{Sato}M. Sato, Y. Kobayashi, S. C. Lee, H. Takahashi, E. Satomi, and Y. Miura, J. Phys. Soc. Jpn. {\bf79}, 014710 (2010).
\bibitem{LiYK}Y. K. Li, X. Lin, Q. Tao, C. Wang, T. Zhou, L. J. Li, Q. B. Wang, M. He, G. H. Cao, and Z. A. Xu, New J. Phys. {\bf11}, 053008 (2009).
\bibitem{FeCuSeCava} A. J. Williams, T. M. McQueen, V. Ksenofontov, C. Felser, and R. J. Cava, J. Phys.: Condens. Matter {\bf21}, 305701 (2009).
\bibitem{LiYK2}Y. K. Li, J. Tong, Q. Tao, C. M. Feng, G. H. Cao, Z. A. Xu, W. Q. Chen, and F. C. Zhang, New J. Phys. {\bf12}, 083008 (2010).
\bibitem{GuoYF}Y. F. Guo, Y. G. Shi, S. Yu, A. A. Belik, Y. Matsushita, M. Tanaka, Y. Katsuya, K. Kobayashi, I. Nowik, I. Felner, V. P. S. Awana, K. Yamaura, E. T. Muromachi, Phys. Rev. B {\bf82}, 054506 (2010).
\bibitem{Chengpeng}P. Cheng, B. Shen, J. P. Hu, and H. H. Wen, Phys. Rev. B {\bf81}, 174529 (2010).
\bibitem{Italy}M. Tropeano, M. R. Cimberle, C. Ferdeghini, G. Lamura, A. Martinelli, A. Palenzona, I. Pallecchi, A. Sala, I. Sheikin, F. Bernardini, M. Monni, S. Massidda, and M. Putti, Phys. Rev. B {\bf81}, 184504 (2010).
\bibitem{GuoYF2}J. Li, Y. F. Guo, S. B. Zhang, S. Yu, Y.
Tsujimoto, H. Kontani, K. Yamaura, and E. Takayama-Muromachi,
Phys. Rev. B {\bf84}, 020513(R) (2011).
\bibitem{JLi2012}J. Li, Y. F. Guo, S. B. Zhang, J. Yuan, Y.
Tsujimoto, X. Wang, C. I. Sathish, Y. Sun, S. Yu, W. Yi, K.
Yamaura, E. Takayama-Muromachi, Y. Shirako, M. Akaogi, and H.
Kontani, Phys. Rev. B {\bf85}, 214509 (2012).

\bibitem{Canfield1} P. C. Canfield, S. L. Budko, N. Ni, J. Q. Yan, and A. Kracher, Phys. Rev. B {\bf80}, 060501(R) (2009).
\bibitem{ZhangGM} G. M. Zhang, Y. H. Su, Z. Y. Lu, Z. Y. Weng, D.
H. Lee, and T. Xiang, Europhys. Lett. {\bf86}, 37006 (2009).
\bibitem{FeCuSeCalc}S. Chadov, D. Scharf, G. H. Fecher, C. Felser, L. Zhang, and D. J. Singh, Phys. Rev. B {\bf81}, 104523 (2010).
\bibitem{AG}A. A. Abrikosov and L. P. Gor'kov, Soviet Phys. JETP {\bf12}, 1243 (1961).

\bibitem{XuG}G. Xu, W. Ming, Y. Yao, X. Dai, S. C. Zhang, and Z. Fang, Europhys. Lett. {\bf82}, 67002 (2008).
\bibitem{Hosono}S. Martsuishi, Y. Inoue, T. Nomura, Y. Kamihara, M. Hirano and H. Hosono, New J. Phys. {\bf11}, 025012 (2009).
\bibitem{3dcalcKuwei}T. Berlijin, C. H. Lin, W. Garber, and Wei Ku, Phys.
Rev. Lett. {\bf108}, 207003 (2012).


\bibitem{3dcalc}H. Wadati, I. Elfimov, and G. A. Sawatzky, Phys.
Rev. Lett. {\bf105}, 157004 (2010).

\bibitem{canfieldcalc}M. G. Kim, J. Lamsal, T. W. Heitmann, G. S. Tucker, D. K. Pratt, S. N. Khan, Y. B. Lee, A. Alam, A. Thaler, N. Ni, S. Ran, S. L. Bud'ko, K. J. Marty, M. D. Lumsden, P. C. Canfield, B. N. Harmon, D. D. Johnson, A. Kreyssig, R. J. McQueeney, and A. I.
Goldman, Phys. Rev. Lett. {\bf109}, 167003 (2012).
\bibitem{HirschfeldReview}P. J. Hirschfeld, M. M. Korshunov and I. I. Mazin. Rep. Prog. Phys. {\bf74}, 124508(2011).

\end{thebibliography}
\end{document}